\documentclass[10pt,conference]{IEEEtran}
\IEEEoverridecommandlockouts
\usepackage{cite}
\usepackage{amsmath,amssymb,amsfonts}
\usepackage{algorithmic}
\usepackage{graphicx}
\usepackage{textcomp}
\usepackage{todonotes}
\usepackage{comment}
\usepackage{tabularx, multirow}
\usepackage{hyperref}
\usepackage{tcolorbox}
\usepackage{booktabs}
\usepackage{enumitem}

\def\BibTeX{{\rm B\kern-.05em{\sc i\kern-.025em b}\kern-.08emT\kern-.1667em\lower.7ex\hbox{E}\kern-.125emX}}

\title{Preprocessing is All You Need: Boosting the Performance of Log Parsers With a General Preprocessing Framework}

\author{\IEEEauthorblockN{Qiaolin Qin, Roozbeh Aghili, Heng Li*\thanks{*Corresponding author.}, Ettore Merlo}
\IEEEauthorblockA{
Polytechnique Montreal, Montreal, Canada \\
\{qiaolin.qin, heng.li, roozbeh.aghili, ettore.merlo\}@polymtl.ca
}
}
\newtcolorbox{boxD}{
    boxsep=0pt,
    left=4pt,right=4pt,top=4pt,bottom=4pt,
    colback = {white}, 
    colframe = {black}, 
    boxrule = 0pt, 
    toprule = 2pt, 
    bottomrule = 2pt 
}

\begin{document}

\maketitle

\begin{abstract}
Log parsing has been a long-studied area in software engineering due to its importance in identifying dynamic variables and constructing log templates. Prior work has proposed many statistic-based log parsers (e.g., Drain), which are highly efficient; they, unfortunately, met the bottleneck of parsing performance in comparison to semantic-based log parsers, which require labeling and more computational resources. Meanwhile, we noticed that previous studies mainly focused on parsing and often treated preprocessing as an \textit{ad hoc} step (e.g., masking numbers). However, we argue that both preprocessing and parsing are essential for log parsers to identify dynamic variables: the lack of understanding of preprocessing may hinder the optimal use of parsers and future research. Therefore, our work studied existing log preprocessing approaches based on Loghub, a popular log parsing benchmark. We developed a general preprocessing framework with our findings and evaluated its impact on existing parsers. Our experiments show that the preprocessing framework significantly boosts the performance of four state-of-the-art statistic-based parsers. Drain, the best statistic-based parser, obtained improvements across all four parsing metrics (e.g., F1 score of template accuracy, FTA, increased by 108.9\%). Compared to semantic-based parsers, it achieved a 28.3\% improvement in grouping accuracy (GA), 38.1\% in FGA, and an 18.6\% increase in FTA. Our work pioneers log preprocessing and provides a generalizable framework to enhance log parsing.

\end{abstract}

\section{Introduction}
\label{sec:introductions}

Logs document software behaviors and are thus powerful in software maintenance. Using the run-time information in log files, researchers can take actions in analyzing and optimizing the systems, such as comprehending the program~\cite{messaoudi2021log, nagaraj2012structured, kuttal2011history, gadler2017mining} and detecting anomalous events~\cite{shin2021theoretical, yang2021semi, zhang2019robust, zhao2021empirical}. The models for log analysis and anomaly detection often require structured inputs such as an event list, while log messages are usually unstructured natural language sentences~\cite{he2017drain}. To bridge the gap between data and models and extract essential static information, log parsers convert the unstructured raw logs to structured formats by masking the variables and extracting event templates. 

Given the importance of log parsers to various downstream tasks, researchers have widely studied log parsing. 
To understand the pros and cons of existing log parsing techniques, a recent study reported the performances of 15 open-sourced log parsers on 14 different log datasets~\cite{jiang2024large}. Among the 15 tools, 13 log parsers are statistic-based (i.e., frequency-based, similarity-based, or heuristic-based), while the remaining 2 log parsers are semantic-based, which use deep learning models for parsing. 
The result revealed that the semantic-based log parsers can obtain higher PA (parsing accuracy) and FTA (F1-score of Template Accuracy) but lower GA (grouping accuracy) and FGA (F1-score of Group Accuracy). The high parsing accuracies of semantic-based log parsers are due to their ability to accurately identify variables. However, the drawbacks of the semantic-based parsers (i.e., UniParser~\cite{liu2022uniparser} and LogPPT~\cite{le2023log}) are also obvious: both tools require more compute resources to run efficiently, and they also require a manual labeling process for model training. 

Conversely, statistic-based log parsers can run label-free and do not demand expensive computing resources such as GPUs. In the life cycle of a statistic-based log parser, log messages often undergo two stages of variable identification: preprocessing and parsing. While many studies focus on enhancing the parsing techniques, detailed guidance on preprocessing is not yet established. Previous studies depict the preprocessing stage as a ``variable-to-placeholder'' substitution process using domain knowledge regular expressions (i.e., regex). The predefined standard regexes differ among datasets (e.g., some datasets only substitute IPv4, while others may also extract time and hexadecimal numbers). The lack of systematical guidance in log preprocessing could confuse researchers and practitioners in dealing with their proprietary, unlabeled log files: \textbf{what variables should we identify before parsing logs?}

To discover the variables to be identified in preprocessing, we shall first understand the characteristics and types of variables. Leveraging the log messages from a widely-used benchmark, Loghub-2k~\cite{zhu2023loghub}, 
Li \textit{et al.}~\cite{li2023did} categorized and studied the words containing variables. However, the so-defined ``words'' (i.e., words split based on blank spaces) may contain static messages or multiple variables. For example, ``type=40031,4508.196000000001,old=4381.818'', contains two different variables: ``40031,4508.196000000001'' and ``4381.818'' and the static parts ``type='' and ``,old=''. Existing labels provided in the study are not sufficient to understand the variable features on a fine-grained level. 

In order to provide detailed guidance on log preprocessing, we also used Loghub-2k to extract all the variables in log messages with the ground truth templates. We then leveraged the default preprocessing regexes for all the datasets to discover the portion of variables that these regexes can match. Further, we analyze the variables that cannot be matched by the provided regexes and develop a new preprocessing framework. We then compare the performance differences of the log parsers before and after using our preprocessing framework. 
Specifically, our study investigates and answers the following research questions:
\begin{itemize}[leftmargin=*]
    \item \textbf{RQ1: What portion of 
    variables can be matched with default regexes in the preprocessing stage? } We collect the variables in each log message using the ground truth template in Loghub-2k, a popular log parsing benchmark with 2,000 log messages for different systems, and gather the default regexes for preprocessing in Loghub. We then analyze how well the domain-knowledge regexes or their combinations can match the variables. 
    \item \textbf{RQ2: What are the characteristics of the unmatched variables?} We categorize the variables not matched by all default regexes in RQ1. The variables are categorized based on their types and semantics. We discuss the detection of variables based on their categories. 
    \item \textbf{RQ3: What is the performance of the preprocessing framework when combined with parsers?} Based on our observations from RQ1 and RQ2, 
    we build a general preprocessing framework. We then combine the framework with four state-of-the-art statistic-based log parsers and compare the performance difference on Loghub 2.0 (an extended version of the log parsing benchmark provided by LogHub~\cite{jiang2024large}\footnote{As explained in \ref{sec:datasets}, we use the extended version of the benchmark to better evaluate the generalizability of our approach.}) 
    before and after using the framework. We also compare the performance of the optimal parser (when combined with our preprocessing framework) with semantic-based parsers. 
    \item \textbf{RQ4: How does the processing framework influence the parsing effectiveness on different log subgroups?} To understand the impact of the preprocessing framework on a finer granularity, we split the log messages according to their different template characteristics, then study the effectiveness impact on log subgroups with different template frequencies and different numbers of variables. 
\end{itemize}

Our work pioneered in improving and studying the impact of log preprocessing; the result suggests that our preprocessing framework brings benefit to all four static-based log parsers examined. We also provide the highly customizable framework and the experimental results in our replication package\footnote{\url{https://github.com/mooselab/Preprocessing-is-All-You-Need}}. 

The remainder of our study is organized as follows. We first motivate our study on log preprocessing in Sec~\ref{sec:motivation}. We then introduce our experiment design in Sec~\ref{sec:study_design}, including the experiment process, the dataset, evaluation metrics, and parsers used throughout this study. We answer the three research questions in Sec~\ref{sec:study_results}. Sec~\ref{sec:threats_to_validity} discusses the validity threats we may encounter in the research. We discuss the previous research related to our study in Sec~\ref{sec:related_work}. We provide the conclusions on our findings in Sec~\ref{sec:conclusion}. 


\section{Motivation}
\label{sec:motivation}

\subsection{The Undesirable Performance Trade-offs in Log Parsers}
\label{sec:statistic_bottleneck}
Given its high importance in log analysis, researchers have studied log parsing for years. Existing log parsers can be categorized as statistic-based and semantic-based~\cite{jiang2024large}. Semantic-based parsers leverage deep learning models to learn the variable patterns with a set of labeled data, classify whether a token is variable, and construct the template based on the predictions~\cite{li2023did, liu2022uniparser, le2023log}. Unlike semantic-based parsers, statistic-based parsers such as LFA~\cite{nagappan2010abstracting} do not require the labor-demanding data labeling and model training but analyze log messages with information such as token frequencies.

According to the results reported by Jiang \textit{et al.}~\cite{jiang2024large}, the highest average parsing accuracy for statistic-based parsers on the Loghub 2.0 dataset is obtained by Drain~\cite{he2017drain}, with a value of 0.470. On the other hand, the semantic-based parser, LogPPT~\cite{le2023log} reached an average parsing accuracy of 0.760, the highest among all the 15 tested parsers. On the scale of average grouping accuracy (GA), Drain reached 0.840, while LogPPT only obtained 0.560. The result suggested an inevitable trade-off in PA and GA when selecting log parsers: the low PA of statistic-based log parsers stands as an obstacle in tasks such as anomaly detection with parameter values~\cite{du2017deeplog}, while the low GA of semantic-based log parsers can lead to problems in monitor occurrence patterns~\cite{khan2022guidelines}. However, it is always desirable to have a log parser with a balanced yet satisfactory performance that can be applied to different downstream tasks; it is suggested to combine semantical and statistical information in parsing~\cite{jiang2024large}. 

We select statistic-based log parsers as our optimization objective in this study because they can efficiently parse large-sized logs even without GPU acceleration~\cite{jiang2024large}. The requirement for data labels also increases the labor cost. Moreover, given the complexity of deep learning models, it is more difficult to interpret and manage the knowledge extracted from the training data. In comparison, statistic-based log parsers are less expensive and more interpretable to use and enhance since the preprocessing step allows explicit knowledge input~\cite{rudin2019stop}. 

\subsection{The Hidden Power in Log Preprocessing}
\label{sec:hidden_power}


Recall that statistic-based parsers usually have two stages in log parsing: \textbf{preprocessing} and \textbf{parsing}. Both steps identify variables in log messages, and the discovered variables in the two stages together contribute to the final construction of templates. During the preprocessing stage, a part of the variables is masked based on domain knowledge (e.g., IPv4, block ID) with placeholders (i.e., \textless{}*\textgreater{}). The domain knowledge consists of explicit variable features summarized by human users and encoded as regexes for variable matching. The knowledge also suggests the semantics hidden behind each matched variable. By correctly leveraging this process, we can potentially suppress the challenge of rare message types or frequently appearing variables that may be mistaken as static text by replacing more variables in the first place. Moreover, the semantical inputs in the preprocessing stage are fully interpretable and manageable compared to the word embeddings learned in semantic-based models. Hence, it is more feasible for users to maintain and monitor the process~\cite{rudin2019stop}. 

However, the pros and cons of the log preprocessing remain unrevealed. Researchers are reluctant to refine the preprocessing process because it has been long believed that the regexes used in the preprocessing stage vary largely across systems. Thus, much domain knowledge and effort are required. To our knowledge, no study reports the impact existing preprocessing processes have on log parsing. Therefore, the effectiveness and whether more effort is demanded for preprocessing remains to be questionable. To remedy this gap in log parsing, our study analyzes log files across 14 different systems and investigates the effectiveness of the existing preprocessing method. 

\subsection{The Problems in the Existing Preprocessing Process}
\label{sec:preprocessing_problems}

Although the parsing process has been widely discussed, and many 
parsers have been established, we lack a thorough study on the effectiveness and impact of the preprocessing stage. Apart from the lack of quantitative analysis mentioned in Sec.~\ref{sec:hidden_power}, the guidelines (e.g., what domain knowledge we should input) of log preprocessing remain to be discussed. 

The scarcity of guidelines hinders researchers and practitioners from preprocessing their proprietary log files without ground truths. For instance, in the default preprocessing function provided by Loghub, paths are masked for OpenStack but not for BGL logs, whereas both log files contain path variables. The unexplained difference in knowledge input can confuse future users when configuring their preprocessing function. 

Moreover, the replacement of variables can be in various formats and is sometimes not aligned with the desired convention, 
causing an inevitable drop in parsing accuracy. For example, an IPv4 address is expected to be converted to ``\textless{}*\textgreater{}:\textless{}*\textgreater{}'' with the Loghub-2k ground truth, but is then converted to ``\textless{}*\textgreater{}'' in the Loghub 2.0 version~\cite{jiang2024large}. The refinement in the newer version does not necessarily indicate that the former mask is wrong: both masks can aid the variable extraction and analysis process but at different granularities. For example, an IPv4 with port (e.g., 10.251.111.130:49851) can be treated as one variable and converted into a single placeholder, while it can be understood on a finer granularity with ``\textless{}*\textgreater{}:\textless{}*\textgreater{}'', which treats the address and the port as two variables. The unified mask may hinder users from extracting their variables on a customized granularity. 
In this study, we aim to provide a guideline for preprocessing log messages based on a thorough analysis of log files across 14 different systems. We then refine the preprocessing regex set and develop a general preprocessing framework based on the findings. Further, we uncover the performance enhancement before and after deploying the new preprocessing framework. 
\section{Experiment Design}
\label{sec:study_design}


\subsection{Overview}
\label{sec:overview}
\begin{figure}[!t]
    \centering
    \includegraphics[width=\columnwidth]{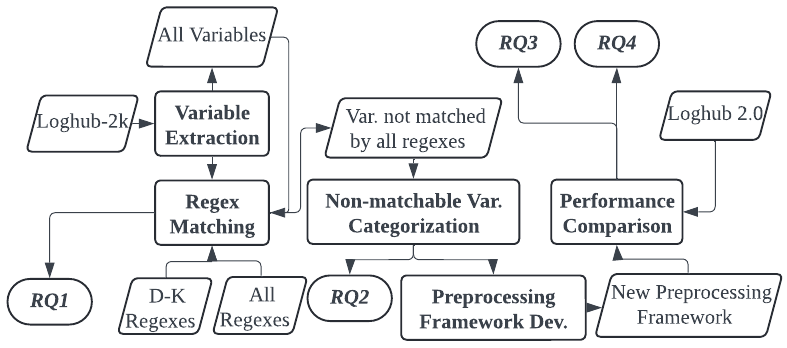}
    \caption{The overview of our study design. D-K stands for domain knowledge.}
    \label{fig:overview}
\end{figure}

Our study is based on the log parsers implementations in Loghub 2.0 and log data provided in both Loghub-2k and Loghub 2.0. 
All the experimental steps are executed on a Mac Mini with an M2 chip and 16GB memory. The study overview is shown in Fig~\ref{fig:overview}. 

\begin{table}[]
\centering
\caption{Two variables are extracted in the log message based on the ground truth template. }
\label{tab:variable_extraction}
\resizebox{\columnwidth}{!}{%
\begin{tabular}{|l|l|}
\hline
\textbf{Message}   & jk2\_init() Found child 8766 in scoreboard slot 12                                             \\ \hline
\textbf{Template}  & jk2\_init() Found child \textless{}*\textgreater in scoreboard slot \textless{}*\textgreater{} \\ \hline
\textbf{Variables} & 8766, 12                                                                                       \\ \hline
\end{tabular}%
}
\end{table}

\subsection{Datasets}
\label{sec:datasets}
We aim to fully understand the characteristics of the variables by studying all the variables in Loghub-2k\footnote{\url{https://github.com/logpai/loghub-2.0/tree/main/2k_dataset}} and answer the first two research questions. The dataset consists of log files and extracted templates from 14 different systems with the updated templates by the original authors~\cite{jiang2024large}; the remaining two systems (i.e., Android and Windows) without template updates are not used. The type of systems includes distributed systems (i.e., Hadoop, HDFS, OpenStack, Spark, Zookeeper), supercomputer systems (i.e., BGL, HPC, Thunderbird), operating systems (i.e., Linux, Mac), Server applications (i.e., Apache, OpenSSH), and standalone software (i.e., HealthApp, Proxifier). Each log file consists of 2,000 sampled log messages with ground truth templates, and the number of templates for each dataset varies from 6 to 341. 

To study the impact of our preprocessing framework on real-world logs, we used the Loghub 2.0 log parsing benchmark, a significantly extended version of Loghub-2k, with a much larger number (i.e., 23,921 to 16,601,745) of annotated log messages and more templates (i.e., 11 to 1,241 templates) for each log dataset~\cite{jiang2024large}. 
Deriving our log preprocessing framework using a smaller benchmark (i.e., Loghub-2k) and evaluating its performance using an extended version (i.e., Loghub 2.0) helps in evaluating the generalizability of our approach. The attributes of Loghub 2.0 made it more accurate in reflecting the real-world usage of parsers. Hence, we used Loghub 2.0 in the last two research questions to better understand the impact of our approach on log parsing performance. 

\subsection{Parser Selection}
\label{sec:parser_selection}
As introduced in Sec.~\ref{sec:introductions}, log parsers can be categorized into statistic-based and semantic-based. Powered by deep learning models, semantic-based log parsers eliminated the need for preprocessing~\cite{li2023did, liu2022uniparser, le2023log} while requiring labels and facing the scalability issue. On the other hand, statistic-based parsers do not need labels and can be more efficient, but often require the preprocessing stage before parsing. According to Jiang \textit{et al.}, only four statistic-based parsers (i.e., Drain~\cite{he2017drain}, IPLoM~\cite{makanju2009clustering}, LFA~\cite{nagappan2010abstracting}, and LogCluster~\cite{vaarandi2015logcluster}) succeeded in parsing all the full-sized log files in a reasonable time window~\cite{jiang2024large}. Therefore, we considered these four log parsers and evaluated our proposed preprocessing framework's impact on their parsing performance. 
The parser with the optimal performance with the help of our preprocessing framework will then be compared with the semantic-based parsers.

\subsection{Experiment Process}
\label{sec:experiment_process}

As shown in Fig~\ref{fig:overview}, our experiments include five main steps, detailed below, to answer the four research questions introduced in Sec.~\ref{sec:introductions}. The specific analysis for answering each individual research is presented in Sec.~\ref{sec:study_results}.

    \noindent \textbf{1) Variable Extraction. }Loghub-2k provides a list of variables for each log message. However, we observed that some extracted variables do not correspond to the  template (e.g., a single variable ``2275'' is documented as four variables ``2'', ``2'', ``7'', ``5''). Therefore, we re-extracted the variables in all log messages in the Loghub-2k dataset with their corrected version of ground truth templates. Table~\ref{tab:variable_extraction} illustrates the process of variable extraction from log messages: ``8766'' and ``12'' are captured using the placeholders (i.e., \textless{}*\textgreater{}). 
    
    \noindent\textbf{2) Regular Expression Matching. } The Loghub benchmark provides a default set of regular expressions, based on domain knowledge, for each log dataset (i.e., the domain-knowledge regexes). We obtain all the default regexes for preprocessing from the Loghub repository. We de-duplicated and categorized the regular expressions as shown in Table~\ref{tab:default_regexes} 
    (e.g., ``(\textbackslash{}d+\textbackslash{}.)\{3\}\textbackslash{}d+'' is a regex for IPv4). We applied 1) only domain-knowledge regexes and 2) all the default regexes after deduplication (10 regexes in total) to identify variables in the log messages. 2 regexes (i.e., block identifier and core identifier) are system-specific, while the remaining 8 regexes are generalizable across systems (i.e., the regexes can be applied to different systems). We compute the precision and recall of both regex sets for matching the variables and collect the variables not identified by all 10 regexes. We aim to answer RQ1 and prepare the variables to be analyzed in RQ2. 
    
    \noindent\textbf{3) Non-matchable Variable Categorization. }We analyze the non-matchable variables obtained from the last step. The variables are categorized according to their semantics and formats (e.g., MAC address). 
    We aim to identify \textbf{generalizable} patterns–those that apply across different systems, thus we are not interested in system-specific variables.
    For example, a variable containing an IPv6 address will be labeled with ``IPv6'', but a customized user name (e.g., admin-1) will be labeled as ``system-specific'', given we cannot summarize a generalizable pattern for such personalized names. We randomly sampled 2,000 variables from all the 23,376 unmatched variables in the combined datasets, which is sufficient to reach a sampling confidence level of 95\% with an error margin of 2\%. Two authors first independently labelled the 2,000 sampled variables, then discussed the range of categories of the variables (\textit{a.k.a}, the codebook), to reach a common understanding of the categories. Further, the two authors independently labelled all the variables with the established categories and Cohen's Kappa was used to measure the agreement ratio. The labels with disagreement were further discussed to reach a consensus. We answer RQ2 with the final labels. 

    \noindent\textbf{4) Preprocessing Framework Development. }Based on the categorization result, we identify the type of generalizable variables ignored in the default preprocessing process. We then refined or enriched the regexes for each of the previously ignored but generalizable variable patterns (e.g., IPv6). 
    

    \noindent\textbf{5) Performance Comparison. }We compare the performance of the log parsers between using the new preprocessing framework and using the default preprocessing, to answer RQ3. We use the metrics described in Sec. \ref{sec:evaluation_metrics} to evaluate the parser's performance. 
    Further, we group the log templates according to their characteristics (i.e., complexity and frequency) and discuss the preprocessing framework's performance impact on different groups, to answer RQ4. 
    

\subsection{Evaluation Metrics}
\label{sec:evaluation_metrics}
Several metrics have been established to evaluate the log parsing quality; the representative metrics are grouping accuracy (GA) and parsing accuracy (PA). Grouping accuracy and its variants evaluate whether log messages sharing the same template also share the template in the ground truth~\cite{zhu2019tools}; the parsing accuracy family estimates the number of templates that are identical to the oracle~\cite{dai2020logram}. According to Khan \textit{et al.}~\cite{khan2022guidelines}, both metrics should be considered for comprehensive analyses. Therefore, as introduced below, we leverage the four metrics used in a large-scale log parser evaluation study~\cite{jiang2024large}. 

\noindent\underline{\textbf{Grouping Accuracy (GA).}} GA calculates the portion of correctly grouped log messages in the dataset by the log parser~\cite{zhu2019tools}. A correct grouping indicates that log messages with the same template in ground truth share the same group. 

\noindent\underline{\textbf{Parsing Accuracy (PA)}} PA calculate the portion of correctly parsed templates~\cite{dai2020logram}. A log is correctly parsed if only all the variables are masked and the static components remain untouched. It assesses the parser's ability to detect and mask the variables in log messages. 

\noindent\underline{\textbf{F1-score of Group Accuracy (FGA)}} FGA is a variant of GA, which considers the portion of correctly grouped templates~\cite{jiang2024large}. FGA is the harmonic mean of the precision and recall of group accuracy. This metric focuses on reporting the grouping performance on the template level. Thus, it provides more weight to the infrequent log templates (e.g., error reports) than the standard GA metric. 
\noindent\underline{\textbf{F1-score of Template Accuracy (FTA)}} FTA considers the number of correctly identified templates~\cite{khan2022guidelines}. The definition of correct identification includes: 1) all the tokens in the template are the same with the ground truth; 2) the log messages corresponding to the template should be the same with the ground truth. FTA is the harmonic mean of the precision and recall of template accuracy. This metric evaluates the parser's ability to identify all variables in each log message. 


\section{Experiment Results}
\label{sec:study_results}
In this section, we report our results obtained through the experimental process described in Sec~\ref{sec:study_design} and answer our research questions introduced in Sec~\ref{sec:introductions}.

\begin{table}[]
\centering
\caption{The default regexes provided in Loghub, along with their semantics. The semantics in italics are system-specific, while the others are generalizable across systems.}
\label{tab:default_regexes}
\resizebox{\columnwidth}{!}{%
\begin{tabular}{l|l}
\hline
\textbf{Regex}                                                                         & \textbf{Semantic}      \\ \hline
\textbackslash{}b(\textbackslash{}-?\textbackslash{}+?\textbackslash{}d+)\textbackslash{}b$|$\textbackslash{}b0{[}Xx{]}{[}a-fA-F\textbackslash{}d{]}+\textbackslash{}b$|$\textbackslash{}b{[}a-fA-F\textbackslash{}d{]}\{4,\}\textbackslash{}b &
  Hexademical/Integer \\ \hline
\textless{}\textbackslash{}d+\textbackslash{}ssec                                      & Time duration          \\ \hline
blk\_-?\textbackslash{}d+                                                              & \textit{Block identifier}       \\ \hline
(/$|$)(\textbackslash{}d+\textbackslash{}.)\{3\}\textbackslash{}d+ & IPv4 \\ \hline
\textbackslash{}b{[}KGTM{]}?B\textbackslash{}b                                         & Memory size unit       \\ \hline
({[}\textbackslash{}w-{]}+\textbackslash{}.)\{2,\}{[}\textbackslash{}w-{]}+(:\textbackslash{}d+)?            & Package name or domain           \\ \hline
core\textbackslash{}.\textbackslash{}d+                                                & \textit{Core identifier}        \\ \hline
=\textbackslash{}d+                                                                    & Assigned Integer       \\ \hline
\textbackslash{}d\{2\}:\textbackslash{}d\{2\}(:\textbackslash{}d\{2\})*                & Time                   \\ \hline
(/.+?\textbackslash{}s$|$(/[\textbackslash{}w-]+)+)                                                               & Path                   \\ \hline
\end{tabular}%
}
\end{table}

\begin{table}[]
\centering
\caption{The precision (P) and recall (R) of variable matching by the domain-knowledge regexes (default regexes for each log file) or all the regexes combined. }
\label{tab:matching_statistics}
\resizebox{0.85\columnwidth}{!}{%
\begin{tabular}{l|ll|ll}
\hline
\multirow{2}{*}{\textbf{Dataset}} & \multicolumn{2}{l|}{\textbf{Domain Knowledge}} & \multicolumn{2}{l}{\textbf{All Regexes}} \\ \cline{2-5} 
                     & \multicolumn{1}{l|}{P}     & R     & \multicolumn{1}{l|}{P}     & R     \\ \hline
\textbf{HDFS}        & \multicolumn{1}{l|}{0.547} & 0.349 & \multicolumn{1}{l|}{0.834} & 0.942 \\ \hline
\textbf{Hadoop}      & \multicolumn{1}{l|}{0.998} & 0.086 & \multicolumn{1}{l|}{0.378} & 0.378 \\ \hline
\textbf{Spark}       & \multicolumn{1}{l|}{0.000} & 0.000 & \multicolumn{1}{l|}{0.427} & 0.540 \\ \hline
\textbf{Zookeeper}   & \multicolumn{1}{l|}{0.000} & 0.000 & \multicolumn{1}{l|}{0.979} & 0.973 \\ \hline
\textbf{BGL}         & \multicolumn{1}{l|}{0.000} & 0.000 & \multicolumn{1}{l|}{0.923} & 0.947 \\ \hline
\textbf{HPC}         & \multicolumn{1}{l|}{0.000} & 0.000 & \multicolumn{1}{l|}{0.838} & 0.795 \\ \hline
\textbf{Thunderbird} & \multicolumn{1}{l|}{1.000} & 0.205 & \multicolumn{1}{l|}{0.834} & 0.565 \\ \hline
\textbf{Linux}       & \multicolumn{1}{l|}{0.564} & 0.178 & \multicolumn{1}{l|}{0.397} & 0.376 \\ \hline
\textbf{HealthApp}   & \multicolumn{1}{l|}{0.000} & 0.000 & \multicolumn{1}{l|}{0.700} & 0.929 \\ \hline
\textbf{Apache}      & \multicolumn{1}{l|}{1.000} & 0.011 & \multicolumn{1}{l|}{0.787} & 0.787 \\ \hline
\textbf{Proxifier}   & \multicolumn{1}{l|}{0.332} & 0.193 & \multicolumn{1}{l|}{0.656} & 0.709 \\ \hline
\textbf{OpenSSH}     & \multicolumn{1}{l|}{0.994} & 0.346 & \multicolumn{1}{l|}{0.874} & 0.725 \\ \hline
\textbf{OpenStack}   & \multicolumn{1}{l|}{0.201} & 0.495 & \multicolumn{1}{l|}{0.201} & 0.473 \\ \hline
\textbf{Mac}         & \multicolumn{1}{l|}{0.249} & 0.026 & \multicolumn{1}{l|}{0.452} & 0.554 \\ \hline
\textbf{Average}     & \multicolumn{1}{l|}{0.421} & 0.135 & \multicolumn{1}{l|}{\textbf{0.691}} & \textbf{0.690} \\ \hline
\end{tabular}%
}
\end{table}

\subsection{RQ1: What portion of variables can be matched with default regexes in the preprocessing stage?}
\label{sec:rq1}

We follow Steps 1) and 2) described in Sec.~\ref{sec:study_design} to identify the variables in the studied log datasets and collect the regexes to match the variables. 
Table~\ref{tab:matching_statistics} reports the precision and recall achieved by using the domain-knowledge regexes only (i.e., the default regexes for each log dataset) or all the default regexes. The precision measures the portion of matched variables that are real variables; the recall measures the portion of all variables that are matched by the regexes.

\setlength{\parskip}{5pt}
\textbf{The domain-knowledge pre-processing regexes can only match an average of 13.5\% variables (i.e., the recall) in the studied log datasets, with an average precision of 42.1\%.} 
With the default regexes, although the matching precision reached 1.000 on two datasets, the low recalls indicate that the provided regexes can only identify a small portion of variables. In fact, most variable matching recalls are lower than 0.3 when matching with only domain-knowledge regexes. We even observed precisions and recalls of 0 on five datasets. Apart from the HealthApp, which does not provide any regex for preprocessing, the 0 recalls on the other four datasets were caused by the inaccuracy of their provided regexes. For example, the regex ``=\textbackslash{}d'' for HPC does not match with variables but matches with the process of value assignment. When used in preprocessing, the whole assignment process will be replaced with a single placeholder, causing the elimination of ``='' symbols and the inaccuracy in parsing. 


\textbf{In contrast, the combined set of pre-processing regexes can match an average of 69.0\% variables, and the average precision increased to 69.1\%.}
We noticed a significant rise in the matching recall when identifying variables with all the regexes provided by Loghub, suggesting the combination can identify more variables. The average recall increased by more than 500\%, suggesting that many variables can be matched by ``non-domain-knowledge'' regexes. In contrast to common belief, these gathered, generalizable regexes matched a high volume of dynamic variables in the majority of log messages: \textbf{8 out of 14 log files reached a recall higher than 0.7}. Meanwhile, using all the regexes for variable detection will increase the average precision by 64\%. However, it is noticeable that the recall dropped with the regex combination for OpenStack. This is because the domain regexes provided for OpenStack leveraged ``((\textbackslash{}d+\textbackslash{}.){3}\textbackslash{}d+,?)+'' to identify multiple IPv4 addresses. We did not take this regex during deduplication because it could lead to the misidentification of IPv4 addresses followed by a comma (e.g., synchronized to 10.100.22.250, stratum 3). The general increment in both recall and precision brought by the combination suggests that different log datasets share generalizable variable patterns, which can be leveraged to improve the preprocessing process.


\begin{boxD}
    During the preprocessing step of log parsing, using merely the domain-specific regexes of an individual log dataset can only match a small portion of variables with a low precision. On the other hand, combining the generalizable regexes of different log datasets can detect significantly more variables (with 5X recall) at a higher precision.  
    Our findings suggest that common variable patterns exist across log files, which may be leveraged to boost log parsing performance. 
\end{boxD}

\subsection{RQ2: What are the characteristics of the unmatched variables?} 
\label{sec:rq2}

We followed Step 3) described in Sec.~\ref{sec:study_design} to categorize all the variables not matched by the regexes in Table~\ref{tab:default_regexes}. 
After obtaining the codebook and labeling all the 23,376 unmatched variables, the Cohen's Kappa of our labels was measured at 0.674 before the final discussion, indicating a substantial agreement~\cite{mchugh2012interrater}. The value increased to 0.989 after the discussion and relabeling. We present the variable categories and their portions among all the unmatchable variables in Table~\ref{tab:nonmathcable_variables}. An example for each category is sampled to facilitate understanding. 

\textbf{We identified 14 categories of unmatched variables, among which 11 categories are generalizable across different systems, covering nearly half of all unmatched ones.}
According to our previous definition, a variable is categorized as \textbf{system-specific} if we cannot summarize a cross-system constant regex based on its semantics (e.g., ``Thunderbird\_A2'' applies only to the Thunderbird system). We also assign a categorize for variables that cannot be matched due to a \textbf{template issue} (e.g., the variable ``32\textgreater{}'' is extracted with template \textless{}memory:\textless{}*\textgreater{}, vCores:\textless{}*\textgreater{}, the real variable should be 32, but was incorrectly extracted due to the unclosed ``\textless{}\textgreater{}''). We label a variable as \textbf{Other} when the variable does not explicitly contain meaningful semantics and cannot be matched using either system-specific or generalizable regexes. The remaining 11 categories (covering 47.3\% variables) can be described with a \textbf{generalizable} regex. We discuss the characteristics and the solutions to these 11 types of variables as follows: 

    \textbf{\underline{Numerical}:} While we only detect digits and hexadecimal variables in the existing regex, there are also float type variables as shown in the example which are missed. Therefore, the regex for numerical variables should be enriched.
    
    \textbf{\underline{Path}:} The path in the example is not matched due to the existence of the ``.'' symbol. The issue can be solved by refining the existing regex for paths. 
    
    \textbf{\underline{Domain}} and \textbf{\underline{IPv4 Related}:} The existing regex follows the variable extraction convention in Loghub-2k and treats domain with ports as one variable. However, the two parts are sometimes treated as separate variables in the template, which violates the convention. Conversely, IPv4 and its port are sometimes combined according to the ground truth, while the convention in Loghub-2k separates them as two variables. This issue is solved in Loghub 2.0 by aligning the masks.
    
    \textbf{\underline{Datetime}:} While the overall format of datetime variables may not be standard as shown in the example variable, the four components in this variable can be masked with generalizable knowledge. To match the free-format date-time information, one can input the list of month and weekday abbreviations for masking; the hexadecimal/integer regex can detect the following numerical components. 
    
    \textbf{\underline{Size}:} Existing regex for memory size only identifies the unit (e.g., KB, GB). However, the memory size and their unit should be treated as one variable according to Loghub-2k's convention. Hence, we shall modify the existing regex.
    
    \textbf{\underline{Package Related}:} The example provided for this category is not matched by the package regex due to the existence of the inner class segmented by the ``\$'' symbol, a convention in Java class reference. We shall enrich the existing package regex to ensure soundness. 
    
    \textbf{\underline{Time Duration}:} The time duration in software is often calculated using units such as ``ms'', ``s'', or ``sec'', while existing regex is limited to a single unit form ``s''.
    
    \textbf{\underline{MAC Address}}, \textbf{\underline{IPv6}}, and \textbf{\underline{URL}}: The three types are usually treated as system variables, and they all can be matched by inputting a constant regex.

To sum up, the 11 variable types can be matched by refining and expanding the existing regex set used for preprocessing. 


\begin{table}[]
\centering
\caption{The categories and portions of variables not matched by the default regexes, with examples.
}
\label{tab:nonmathcable_variables}
\resizebox{\columnwidth}{!}{%
\begin{tabular}{l|l|l}
\hline
\textbf{Category}        & \textbf{Portion} & \textbf{Example}                             \\ \hline
\textbf{System Specific} & 0.272            & Thunderbird\_A2                              \\ \hline
\textbf{Numerical}       & 0.236            & 1499518522.558304                            \\ \hline
\textbf{Template Issue}  & 0.226            & 32\textgreater{}                             \\ \hline
\textbf{Path}            & 0.077            & /etc/httpd/conf/workers2.properties          \\ \hline
\textbf{Domain}          & 0.067            & FNANLI5.fareast.corp.microsoft.com           \\ \hline
\textbf{Datetime}        & 0.039            & Wed Jun 22 13                                   \\ \hline
\textbf{IPv4 Related}    & 0.021            & 203.205.151.204:80                                     \\ \hline
\textbf{Other}           & 0.019            & \textless{}ok\textgreater{}                  \\ \hline
\textbf{Size}            & 0.017            & 64172MB                                      \\ \hline
\textbf{Package Related} & 0.01             & v2.app.launcher.ContainerLauncher\$EventType \\ \hline
\textbf{MAC Address}     & 0.003            & 00:11:43:e3:ba:c3                            \\ \hline
\textbf{Time Duration}   & 0.002            & 10000ms                                      \\ \hline
\textbf{IPv6}            & 0.002            & 2607:f140:6000:8:c6b3:1ff:fecd:467f          \\ \hline
\textbf{URL}             & 0.002            & https://gsp-ssl.ls.apple.com/dispatcher.arpc \\ \hline
\end{tabular}%
}
\end{table}
\begin{boxD}
    Existing non-matchable variables can be categorized into 14 categories, including 11 generalizable categories covering 47.3\% of all the unmatched variables, which can be identified by refining or expanding the existing generalizable processing regexes. 
\end{boxD}

\subsection{RQ3: What is the performance of the preprocessing framework when combined with parsers?}
\label{sec:rq3}
\subsubsection{\underline{General Preprocessing Framework}}
\label{sec:framework_refinement}

According to the findings in RQ2, we recognized several categories of variables that are not captured in the default regexes but can be detected by adding to or modifying their corresponding regexes. 
Provided with this knowledge, we refined and expanded the regexes for masking. The refined set comprises 15 regexes, including those for memory size, date, time, URL, etc. As introduced by Jiang \textit{et al.} and also mentioned in Sec.~\ref{sec:preprocessing_problems}, the masking fashion differs in Loghub-2k and Loghub 2.0. Our framework supports the customizable assignment of mask formats to each regex (e.g., \textless{}*\textgreater{}:\textless{}*\textgreater{}:\textless{}*\textgreater{} for time variables). For evaluation, we used the default masking technique of converting all the variables to a single placeholder \textless{}*\textgreater{}. The regexes shall also follow a certain order to preprocess the log messages, especially when two regexes share similar parts. For instance, if the times are masked before a MAC address ``00:00:00:12:34:56'', the address will be replaced as ``\textless{}*\textgreater{}:\textless{}*\textgreater{}'', while the ground truth of MAC address replacement is ``\textless{}*\textgreater{}''. Therefore, MAC addresses should be masked before time. The detailed regexes, along with their orders and customizable usage introduction can be found in our replication package.


The time required by regular expression matching and substitution sublinearly increases when the number of regexes rises, while some systems do not contain several types of variables. For example, OpenSSH does not have MAC addresses as variables. Therefore, for efficiency, it is necessary to filter the regexes that do not apply to the dataset. However, whether a regex is required for preprocessing remains unclear before log parsing, and it can barely be determined using domain knowledge, as explained in RQ1. To tackle this issue, we estimate whether a regex is applicable using the first $n$ lines of the log messages. This method serves as an evidence-based estimation of the type of variables in a system. We set $n$ to 1,000, 2,000, and 3,000 to calculate the portion of the regex that can be correctly selected using the first $n$ lines. When set to 1,000, the portion is 0.929; when increased to 2,000 and 3,000, the portion is raised to 0.989. This indicates that if a type of variable does not exist in the first 2000 lines, it has a low probability of appearing in the remaining messages in the same system. Thus, we used the first 2,000 message lines in each dataset for the estimation. 


Our framework is developed and evaluated based on Loghub 2.0. The results for each parsing evaluation metric are obtained from executing the script provided in Loghub 2.0~\cite{jiang2024large}. 

\begin{figure*}[!t]
    \centering
    \includegraphics[width=\linewidth]{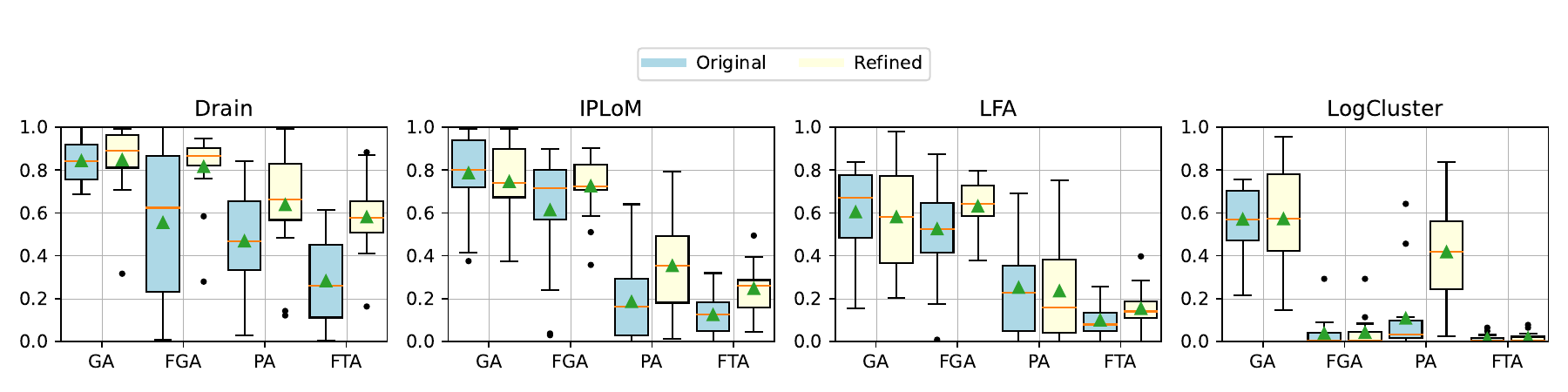}
    \caption{The evaluation results of the four statistic-based parsers. The blue boxes indicate the parsers with the original preprocessing function, while the yellow boxes show the results of parsers with the new preprocessing framework. The red lines show the medians, and the green arrows indicate the means.}
    \label{fig:performance_comparison}
\end{figure*}
\begin{table}[]
\centering
\caption{The change in average GA, FGA, PA, and FTA 
after using our preprocessing framework. 
}
\label{tab:performance_difference_full}
\resizebox{0.9\columnwidth}{!}{%
\begin{tabular}{l|l|l|l|l}
\hline
\textbf{Metric}      & \textbf{Drain} & \textbf{IPLoM} & \textbf{LFA} & \textbf{LogCluster} \\ \hline
\textbf{Average GA}  & +0.8\%       & -10.1\%      & -3.2\%     & -5.6\%            \\ \hline
\textbf{Average FGA} & +48.4\%      & +17.4\%      & +20.8\%    & +2.5\%              \\ \hline
\textbf{Average PA}  & +35.5\%      & +67.4\%      & -5.6\%       & +277.3\%          \\ \hline
\textbf{Average FTA} & +108.9\%     & +95.0\%        & +59.0\%      & +60.0\%             \\ \hline
\end{tabular}%
}
\end{table}

\subsubsection{\underline{Overall Performance Impact on Parsers}}
\label{sec:overall_parsers}
We combine the preprocessing framework with the four parsers. The results obtained are shown in Fig~\ref{fig:performance_comparison} and Table~\ref{tab:performance_difference_full}.  The box plots in Fig~\ref{fig:performance_comparison} show the distribution of the metrics. 
Table~\ref{tab:performance_difference_full} shows the average performance difference in the four metrics.


\textbf{In general, combining our preprocessing framework with statistics-based log parsers can significantly improve their template level accuracy (FGA and FTA).}
Both FGA and FTA improved for all four tools, according to Table~\ref{tab:performance_difference_full}, showing that the preprocessing framework can enhance the grouping and parsing accuracy on the template level. Notably, the average FGA for Drain increased by 48.4\%, and the average FTA increased by 108.9\%. The rise in FGA suggests that our preprocessing framework can also accurately replace the variables in rarer log templates and achieve better grouping performance on the template level, and the increment of FTA indicates that log parsers can benefit from leveraging the rich semantic knowledge input in the preprocessing stage and thus provide more accurate log templates for downstream tasks. 

\textbf{The PA also shows a generally increasing trend with our preprocessing framework.} For example, the average PA of Drain increases by 35.5\%. The only exception is the parser LFA. 
As a frequency-based parser, LFA depends heavily on the similarity of the log messages and the token frequencies in each message. The variable masks created in the preprocessing stage may cause two different, non-common log messages with the same lengths to be closer. The change in preprocessing will then lead to some merging errors: for example, ``[client 59.120.212.70] script not found or unable to stat: /var/www/cgi-bin/awstats'' and ``[client 63.160.17.140] request failed: URI too long (longer than 8190)'' are grouped into one template because their length is the same and both contain two variable masks on the same position (i.e., the second token and the last token). The group, therefore, will treat the majority of the tokens as variables. 


\textbf{In contrast to the general increment trend on FGA, we observed an unstable impact on GA. }The average GA before and after deploying the new framework fluctuates: Drain obtained a slightly higher GA after the combination, while the other three suffered from some extent of decrements. As suggested by Jiang \textit{et al.}~\cite{jiang2008abstracting}, searching variables based on semantics or types may cause ignorance of global information such as statistical frequency and thus may harm the grouping performance. For example, the class ``java.net.ConnectException'' included in Hadoop is regarded as a constant component, when classes are usually treated as a variable type and are masked during preprocessing.  


\begin{table}[]
\centering
\caption{The average performance of Drain before (Original Drain) and after (New-Pre Drain) combined with our preprocessing framework, compared to the performance of the two semantic-based parsers. 
}
\label{tab:performance_difference_st_se}
\resizebox{0.9\columnwidth}{!}{%
\begin{tabular}{l|l|l|l|l}
\hline
\textbf{Avg. Metric}    & \textbf{GA} & \textbf{FGA} & \textbf{PA} & \textbf{FTA} \\ \hline
\textbf{Original Drain}          & 0.843        & 0.554       & 0.468       & 0.282        \\ \hline
\textbf{New-Pre Drain}          & \textbf{0.847}        & \textbf{0.815}       & 0.638       & \textbf{0.581}        \\ \hline
\textbf{UniParser} & 0.660       & 0.500       & 0.680        & 0.260        \\ \hline
\textbf{LogPPT} & 0.560       & 0.590       & \textbf{0.760}        & 0.490        \\ \hline
\textbf{Improvement v.s. Sem.-based} & +28.3\%       & +38.1\%       & -16.1\%        & +18.6\%        \\ \hline
\end{tabular}%
}
\end{table}

\textbf{When combined with our preprocessing framework, Drain obtained the best performance on all four metrics}. Thus, we compared its performance with semantic parsers (i.e., UniParser and LogPPT). Table~\ref{tab:performance_difference_st_se} shows the GA, FGA, PA, and FTA obtained by the updated Drain and the original UniParser or LogPPT. The scores of UniParser and LogPPT are obtained from the effectiveness report by Jiang \textit{et al.}~\cite{jiang2024large}. According to Table~\ref{tab:performance_difference_st_se}, Drain with the new preprocessing framework has a more balanced performance and obtained significantly higher GA (28.3\% improvement), FGA (18.6\% improvement), and FTA (18.6\% improvement) than the optimal performance of the two semantic-based parsers. With a more comprehensive preprocessing framework, rare templates can also be grouped and parsed by Drain with higher accuracy. However, the PA gained by Drain is still slightly lower than LogPPT. This is because some system-specific variables are among the first several tokens, and thus cannot be detected due to Drain's assumption. For instance, ``Kind: YARN\_AM\_RM\_TOKEN'' are the first two tokens in a template of Hadoop and, therefore, are treated as constant elements. However, ``YARN\_AM\_RM\_TOKEN'' is a variable in the ground truth. 

\begin{boxD}
    Combined with our preprocessing framework, the statistic-based log parsers show a significant overall performance improvement, especially in terms of the template-level metrics (FGA and FTA). 
    With the help of the framework, the best statistic-based parser (i.e., Drain) can obtain a more balanced performance and outperform the two semantic-based log parsers regarding GA, FGA, and FTA. 
\end{boxD}

\subsubsection{\underline{Efficiency Impact on Parsers}}
\label{sec:efficiency_impact}

As previously discussed in Sec.~\ref{sec:framework_refinement}, the time cost of the framework increases sublinearly when the number of regexes used for preprocessing rises. The time increment is minor on systems with a few log lines (e.g., Proxifier with $\sim$ 20k lines), but may be significant when dealing with large log files (e.g., Thunderbird, where the log lines are more than 1,500k). Therefore, we filtered the variable types unseen in the first 2,000 log lines.

\begin{figure*}[!t]
    \centering
    \includegraphics[width=0.9\linewidth]{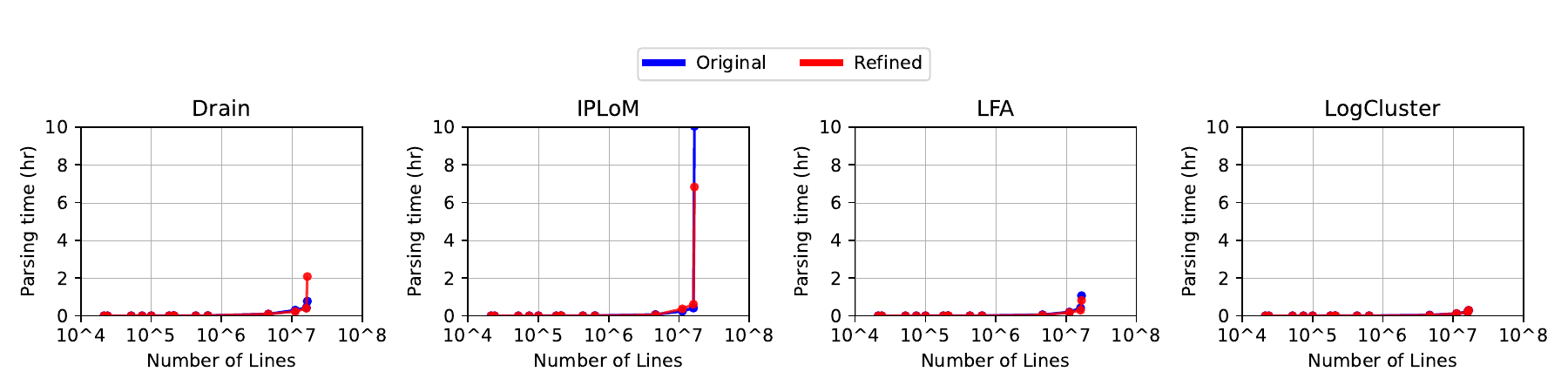}
    \caption{The number of hours required for parsing each log file using different parsers. }
    \label{fig:efficiency}
\end{figure*}

The time consumption of the parsers with and without implementing the new framework on each dataset are shown in Fig.~\ref{fig:efficiency}. The parsers of each version were run one at a time on the same machine. Drain, LFA, and LogCluster succeeded in parsing all the logs in three hours. IPLoM tends to require much more execution time when parsing larger files, but it can still manage to parse the largest file (i.e., Thunderbird) in 12 hours. The refinement of preprocessing has generally reduced the time cost of IPLoM by extracting variables beforehand. On the other hand, we observed a large difference in parsing Thunderbird with Drain: the execution time almost tripled after the preprocessing refinement. The major increment is in the Loghub variable extraction function of the result storage stage, since the preprocessing stage changes some of the tokens in the log message, it could impact the behavior of the extraction function. 

\begin{boxD}
    The increased preprocessing of our framework adds extra computational time, whereas it can also decrease the complexity of the parsing process itself (i.e., fewer variables to be parsed). 
    Therefore, the end-to-end time complexity is not deterministic.
\end{boxD}

\subsection{RQ4: How does the processing framework influence the parsing effectiveness on different log subgroups?}
\label{rq4}

\begin{figure*}[!t]
    \centering
    \includegraphics[width=0.9\linewidth]{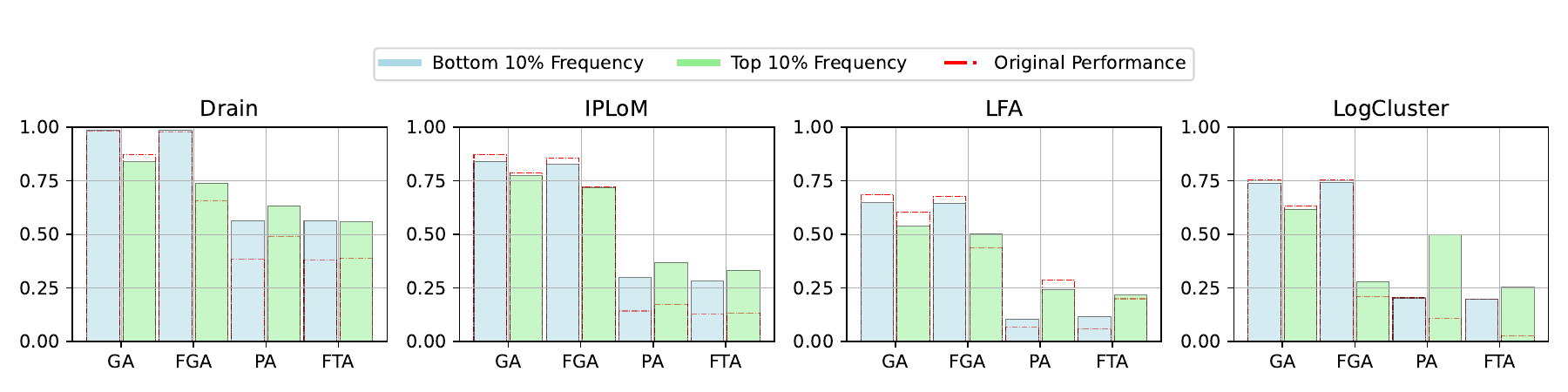}
    \caption{The average evaluation results of log parsers on logs with different frequencies (i.e., the most frequent 10\% and the least frequent 10\%.) The red dot lines illustrate the original results obtained with the previous preprocessing function. }
    \label{fig:template_frequencies}
\end{figure*}
\begin{figure*}[!t]
    \centering
    \includegraphics[width=0.9\linewidth]{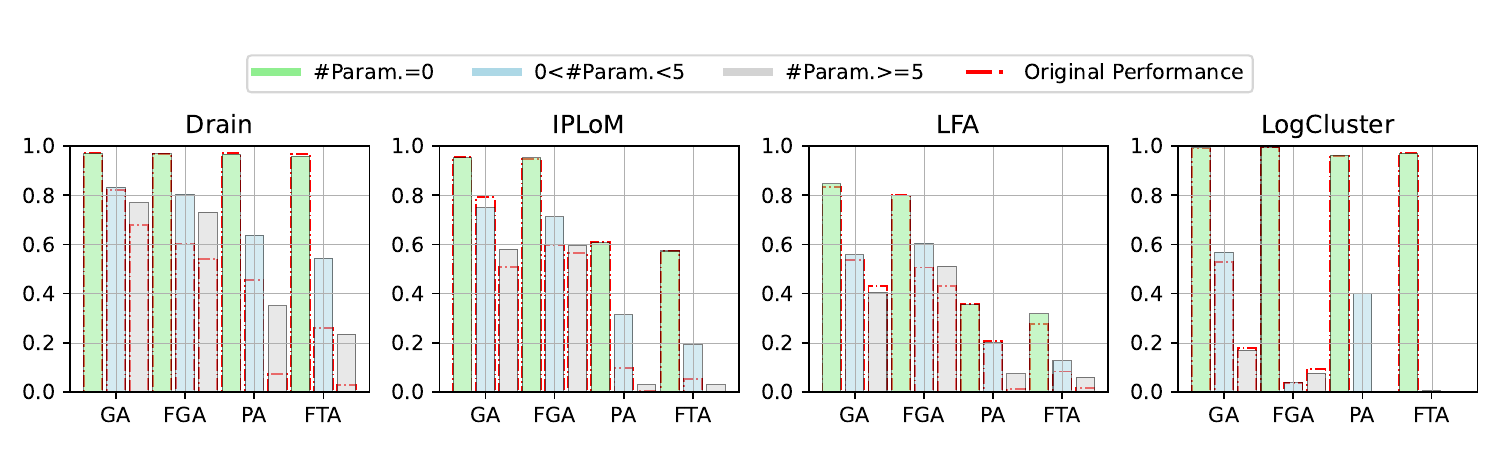}
    \caption{The average evaluation results of log parsers on logs with different numbers of variables. The red dot lines illustrate the original results obtained with the previous preprocessing function. }
    \label{fig:template_complexities}
\end{figure*}

\subsubsection{\underline{Different Template Frequencies}}
We compare the impacts of the preprocessing framework on the parsing performance of the top 10\% most frequent and the least frequent 10\% templates in Fig.~\ref{fig:template_frequencies}. As previously analyzed in Sec.~\ref{sec:overall_parsers}, the additional semantic input provided through the framework may slightly reduce the grouping performance by masking all ``variable-like'' tokens; therefore, the grouping accuracies show a generally decreasing trend among tools except for Drain. 

\textbf{Our preprocessing framework can improve the parsing results for both the most frequent and least frequent templates.}
The majority of PAs and FTAs increased or remained the same value after the modification, especially for the bottom 10\% frequency templates. The FTA for Drain exceeded 0.5 on both the most and least frequent template groups. The results are higher than LogPPT's FTA reported by Jiang \textit{et al.}~\cite{jiang2024large}, showing that Drain with our preprocessing framework is highly effective. Conversely, the PA for LFA on the most frequent template group decreased since the token change in the preprocessing stage influenced the similarity between log messages, as addressed in Sec.~\ref{sec:overall_parsers}. We also observed that the parsing performances for LogCluster did not improve on the least frequent templates due to its assumption. 

\begin{boxD}
    Using the slight decrement of grouping accuracies as a trade, the new preprocessing framework can generally and significantly improve both PA and FTA, regardless of the template frequencies. For example, combined with our framework, Drain obtained a higher FTA for both the most frequent and least frequent templates, outperforming LogPPT. 
\end{boxD}

\subsubsection{\underline{Different Template Complexities}}
Template complexity is defined based on the number of variables in a template~\cite{jiang2024large}. We split the templates into three groups: templates contain no variables (i.e., \#Param.=0), templates containing less than 5 variables (i.e., 0\textless{}\#Param.\textless{}5), and variable-intensive templates (i.e., \#Param.\textgreater{}5). We then report the effectiveness of our preprocessing framework for each group in Fig.~\ref{fig:template_complexities}. 

\textbf{Our preprocessing framework shows a larger benefit when the number of variables in the logs increases.}
When the template have no variables, the new preprocessing framework will not change or only subtly reduce the performance in some cases (e.g., a static token ``PS/2'' is wrongly treated as a variable). This shows that using a wide range of generalizable variable regexes will raise only a few false positives. 

The benefit of our framework amplifies when the template complexity increases. The original parsers can barely identify all the variables in variable-intensive logs because of the large population. The small number of templates in this group also made elevating the quantitative performance hard: 12 datasets have only less than 20 variable-intensive templates, and thus, the parser should be highly precise to obtain high scores. With the new framework, all four metrics were boosted when using Drain to parse the log messages containing variables, especially for the variable-intensive log group. This improvement overcomes the concern of Jiang \textit{et al.}~\cite{jiang2024large}, that the low grouping and parsing accuracies on these templates would lead to distracting parsing errors in applications. 

However, the framework did not largely enhance the variables-intensive log groups' performances for IPLoM. The tool has a delimiter processing stage during log parsing, and symbols such as ``='' are used as delimiters and are removed from the final templates. This assumption would lead to the decrement of PA and FTA. While ``='' often indicates a variable assignment process, it has a higher chance of existing in variable-intensive templates. Given that these symbols are eliminated, the PA and FTA of IPLoM will still be low even if all the variables are accurately and precisely identified. 

\begin{boxD}
    The preprocessing framework is unlikely to cause false positives for templates without variables. On the other hand, it can boost the parsing performances on the variable-intensive templates.
\end{boxD}

\section{Threats to Validity}
\label{sec:threats_to_validity}

\noindent\underline{\textbf{Internal Validity.}} This evaluation of this work relies on the ground truth data provided by Loghub 2.0. However, we noticed there are still some minor issues with the ground truth templates provided in Loghub 2.0. For instance, the angle bracket is not closed in one of the templates for Hadoop and writes as ``\textless{}memory:\textless{}*\textgreater{}, vCores:\textless{}*\textgreater{}''. Based on the template, the second variable would be written as, for example, ``32\textgreater{}'', while the actual variable is ``32''. This type of issue would cause inevitable inaccuracy in log parsing. However, to ensure equity when comparing with precedent studies, we did not modify the ground truths. 

The evaluation of execution time may be impacted by I/O and the CPU usage status. We suppressed the issues by running the original and the refined codes with the same environment and configuration. 

\noindent\underline{\textbf{External Validity.}} We studied the characteristics of the variables and refined the regex set for preprocessing on 14 datasets provided in Loghub, covering 4 types of systems. Although these datasets are wide in variety and are commonly used in log-related studies, some generalizable variables may be unseen in the dataset. These issues can be easily solved by configuring the regex set with the user's domain knowledge when processing their own private systems. 

\noindent\underline{\textbf{Construct Validity.}} 
Our preprocessing framework is refined based on the analyses of the variables, and the categorization may vary among individuals (e.g., ``2005'' may be categorized as either DateTime or numerical) according to different criteria. Hence, the labels for the variables are labeled by two authors independently and discussed to ensure consistency. 
\section{Related Work}
\label{sec:related_work}

\noindent\underline{\textbf{An Overview of Log Parsing Studies.}} 
Log parsing is essential in the field of log analysis, and a lot of log parsers have been established. To understand the pros and cons of the parsing algorithms, Zhu \textit{et al.}~\cite{zhu2019tools} evaluated 13 log parsers on Loghub-2k, a dataset containing 16 different system log files with 2k lines of log messages in each~\cite{zhu2023loghub}. Khan \textit{et al.}~\cite{khan2022guidelines} noticed that previous studies on log parser performances are not aligned in metrics and carried out a study on log parsing evaluation metric guidelines to explain the differences in metrics such as grouping accuracy and parsing accuracy. Following the guidance, Jiang \textit{et al.}~\cite{jiang2024large} comprehensively re-evaluated the existing state-of-the-art log parsers using a refined version of Loghub (i.e., Loghub 2.0). Although previous works studied log parsing in detail, they merely focused on the parsing algorithm but neglected log preprocessing, which is also an essential stage in statistic-based log parsing. Our work aims to remedy this gap by studying the power and providing guidelines on this process. 

\noindent\underline{\textbf{Statistic-Based Log Parsers.}} 
Statistic-based log parsers form templates according to the statistical facts in log messages, such as token frequencies~\cite{jiang2024large}. These parsers can be further categorized as frequency-based, similarity-based, or heuristic-based. Frequency-based parsers such as Logram, distinguish the static and dynamic parts in log messages with their existence frequency~\cite{dai2020logram, vaarandi2003data, nagappan2010abstracting, vaarandi2015logcluster}. Based on different definitions of distances and similarity, similarity-based parsers group log messages with clustering algorithms, and then summarize a template for each group~\cite{fu2009execution, hamooni2016logmine, mizutani2013incremental, shima2016length, tang2011logsig}. Heuristic-based parsers group the log messages and summarize templates based on different heuristic algorithms or data structures such as parsing trees or longest common subsequence-based approach~\cite{du2016spell, he2017drain, makanju2009clustering, messaoudi2018search, yu2023brain}. In the large-scale evaluation by Jiang \textit{et al.}~\cite{jiang2024large}, these parsers share a similar preprocessing stage but vary in parsing algorithms. According to the reported results, statistic-based log parsers can reach high grouping accuracies but are generally low in parsing accuracies. Our work refine the preprocessing process shared by these parsers to break the parsing accuracy bottleneck while maintaining the same level of grouping performance. 


\noindent\underline{\textbf{Semantic-Based Log Parsers.}} 
In contrast to statistic-based log parsers, semantic-based log parsers directly identify the variables based on their semantical features and thus do not require the preprocessing stage~\cite{jiang2024large}. These parsers learn to classify features with a portion of labeled log message and extract the variables in log messages through an inference stage~\cite{liu2022uniparser,le2023log,li2023did}. According to the study by Jiang \textit{et al.}~\cite{jiang2024large}, these parsers require more computational costs and usually have higher parsing accuracy but lower grouping accuracy in comparison to statistic-based parsers. 
\section{Conclusion}
\label{sec:conclusion}
Our work performs a comprehensive study of log preprocessing, a rarely explored but important step in log parsing. 
First, we observe that, by combining the generalizable knowledge in the preprocessing steps of different log datasets, we can significantly enhance the precision and recall of log preprocessing in terms of variable matching.
Through a comprehensive analysis of the existing preprocessing approaches (regexes) and their gap with the variables in the log datasets, we develop a general log preprocessing framework that leverages and improves the generalizable knowledge in the existing preprocessing steps. In general, combining our preprocessing framework with statisict-based log parsers can significantly improve theireffectivenesss. The state-of-the-art statistic-based log parser (i.e., Drain) improved on all four metrics after implementing the new framewor: the FTA noticeably improved by 108.9\%.k It even outperforms semantic-based log parsers in terms of FTA, GA, and FGA. The results revealed the power of log preprocessing and suggested that the semantic input from the preprocessing stage can greatly boost the performance of static-based log parsers. Future research or practices should pay more attention to the preprocessing step to improve their log parsing results.


\bibliographystyle{IEEEtran}
\bibliography{references}
\end{document}